\documentclass{IJCAS}

\usepackage{url}
\usepackage{array,tabularx}
\usepackage{multicol}

\usepackage{amsmath,amsfonts}
\usepackage{algorithmic}
\usepackage{array}
\usepackage{textcomp}
\usepackage{stfloats}
\usepackage{url}
\usepackage{verbatim}
\usepackage{graphicx}
\usepackage{amsthm}
\usepackage{cite}
\usepackage{textcomp}
\usepackage[caption=false, font=footnotesize]{subfig}
\usepackage{enumerate}

\journalvolumn{VV}
\journalnumber{X}
\journalyear{YYYY}
\setarticlestartpagenumber{1}

\allowdisplaybreaks

\begin{document}
\title{Bridging the Sim-to-real Gap:\\A Control Framework for Imitation Learning of Model Predictive Control}

\author{Seungtaek Kim\orcid{0000-0001-8412-9458}, Jonghyup Lee\orcid{0000-0002-1022-5129}, Kyoungseok Han*\orcid{0000-0002-4986-2053}, and Seibum B. Choi*\orcid{0000-0002-8555-4429}}

\begin{abstract}
To address the computational challenges of Model Predictive Control (MPC), recent research has studied using imitation learning to approximate MPC with a computationally efficient Deep Neural Network (DNN). However, this introduces a common issue in learning-based control, the simulation-to-reality (sim-to-real) gap. Inspired by Robust Tube MPC, this study proposes a new control framework that addresses this issue from a control perspective. The framework ensures the DNN operates in the same environment as the source domain, addressing the sim-to-real gap with great data collection efficiency. Moreover, an input refinement governor is introduced to address the DNN's inability to adapt to variations in model parameters, enabling the system to satisfy MPC constraints more robustly under parameter-changing conditions. The proposed framework was validated through two case studies: cart-pole control and vehicle collision avoidance control, which analyzed the principles of the proposed framework in detail and demonstrated its application to a vehicle control case.
\end{abstract}


\maketitle

\makeAuthorInformation{
Manuscript received XX XX, XXXX\\

Seungtaek Kim and Seibum B. Choi (Corresponding Author) are with the Dept. of Mechanical Engineering, Korea Advanced Institute of Science and Technology, Daejeon 34141, South Korea (email: kimst9o9@kaist.ac.kr, sbchoi@kaist.ac.kr). Jonghyup Lee is with the Dept. of Mechanical Systems Engineering, Sookmyung Women's University, Seoul 04310, South Korea (email: jhyuplee@sookmyung.ac.kr). Kyoungseok Han (Co-corresponding Author)  is with the Dept. of Automotive Engineering, Hanyang University, Seoul 04763, South Korea (email: kyoungsh@hanyang.ac.kr).

* Corresponding authors.
}

\runningtitle{2025}{Seungtaek Kim, Jonghyup Lee, Kyoungseok Han, and Seibum B. Choi}{Bridging the Sim-to-real Gap:A Control Framework for Imitation Learning of Model Predictive Control}{xxx}{xxxx}{x}

\section{INTRODUCTION}

Model Predictive Control (MPC) has been widely applied across various control domains due to its ability to compute optimal control inputs while satisfying system constraints. Therefore, MPC has shown strong applicability to constraint-driven control problems such as powertrain transmission control (jerk constraints)~\cite{lee2023real}, vehicle path tracking control (road friction limit constraints)~\cite{lee2022integrated}, and legged robot locomotion (slip motion constraints)~\cite{hong2020real}. However, as system complexity increases, the computational burden of solving the underlying optimization problem also grows significantly, making real-time implementation challenging.

Many studies have focused on reducing its high computational cost. One attractive approach is to approximate MPC with a computationally efficient Deep Neural Network (DNN) using imitation learning \cite{karg2020efficient}. This method involves collecting demonstrations from MPC and training a DNN to imitate its behavior through supervised learning. Studies have shown that this method reduced MPC's computational load by at least 10 to 20 times while maintaining its control performance, verified by gait control for quadrupedal robots \cite{reske2021imitation}, humanoid's bipedal control \cite{vitor2025imitation}, drone  \cite{zhang2016learning}, and vehicle motion control \cite{park2025deep,lee2022real}.

However, applying imitation learning inevitably introduces the simulation-to-reality (sim-to-real) gap, a major issue in the learning-based control area. This gap refers to the disparity between the source domain (where the DNN's training data is collected, e.g., a simulation or lab) and the target domain (where the DNN is applied in the real world). Since collecting MPC demonstrations in the real world is impractical, many studies have instead collected them in a simulated environment, thereby introducing a sim-to-real gap. If this gap is not addressed before applying the DNN to the target domain, the DNN may encounter untrained states, leading to unintended control failures.

Various transfer learning techniques \cite{zhao2020sim} have been developed in learning-based control to address the sim-to-real gap, including domain adaptation, meta learning, and Domain Randomization (DR). Domain adaptation\cite{farahani2021brief,chebotar2019closing} is primarily used in computer vision, aiming to pre-align the target domain with the source domain to minimize the domain gap itself. Meta-learning\cite{clavera2018model} takes a similar perspective but focuses on fast adaptation of the controller, allowing a policy trained in the source domain to quickly adapt to new or unseen dynamics in the target domain.

On the other hand, Domain Randomization (DR) compensates for the sim-to-real gap by introducing random perturbations, such as variations in model parameters and friction \cite{andrychowicz2020learning}, wind disturbances \cite{wada2022sim}, and sensor noise \cite{wang2020reinforcement}, into the source domain so that it encompasses the variability of the target domain. DR has been widely adopted in many control applications as a practical solution for sim-to-real transfer because it does not require any access to the target domain, which is often costly, time-consuming, or even unsafe. However, several drawbacks still remain: (1) determining which factors to randomize and how to do so is challenging, (2) the amount of training data increases drastically as more random factors are considered, (3) the controller becomes excessively conservative to account for all random variations, and (4) despite these efforts, the randomized elements may not fully capture the gap, making the DNN still vulnerable to sim-to-real gap.

Therefore, various studies have proposed advanced DRs to overcome these limitations. A study applied the concept of a bounded-error tube as a criterion for randomizing factors \cite{tsukamoto2021learning}, and, to further improve data-collection efficiency, a method for extracting samples from each vertex or face of a polytope error tube was proposed \cite{tagliabue2024efficient}. However, these approaches still required additional perturbed data and resulted in a conservative control strategy similar to other DR methods. To address parameter changes in the target domain, a parameter-adaptive approximate MPC \cite{hose2024parameter} was proposed, which attaches a linear predictor to the DNN controller to compensate for parameter changes. However, a significant amount of MPC demonstrations was still required to overcome the remaining sim-to-real gap.

This study, on the other hand, proposes a new control framework that overcomes the limitations (1) - (4) of existing DR methods. While previous approaches focus on reducing the sim-to-real gap, our method forces the DNN to operate within the source domain. Inspired by the Robust Tube MPC (RTMPC) structure, the proposed framework configures the approximate DNN as a nominal controller that operates only on the nominal model, ensuring it remains within the source domain state distribution. This prevents instability caused by the sim-to-real gap, as the DNN is prevented from being exposed to an untrained state. In the proposed structure, as the DNN only controls the nominal model, acting as a feedforward controller, an additional stable ancillary controller is introduced to guide the actual plant to follow the nominal model. 

Furthermore, inspired by the parameter governor concept \cite{kolmanovsky2006parameter}, an input refinement governor, an add-on mechanism, refines the control inputs from both controllers, allowing the overall controller to adapt to changes in plant model parameters that the DNN alone cannot handle. It enforces MPC constraints and allows a less conservative controller to be imitated. To validate the proposed framework, two case studies were conducted: cart-pole and vehicle collision avoidance systems, which provide a detailed explanation of its components and demonstrate its superior performance in handling the sim-to-real gap compared to existing DR methods.

The key contributions of the proposed framework are summarized as follows:  
\begin{enumerate}[{1)}]
\item It improves data efficiency for addressing the sim-to-real gap compared to other DR methods.  
\item A newly proposed source domain is inherently defined during MPC design, reducing the need to create a high-fidelity source domain.  
\item The proposed framework enables the imitation of a less conservative MPC handling the variations in the plant's model parameters.
\end{enumerate}

The remainder of this paper is organized as follows: Section \uppercase\expandafter{\romannumeral2} presents the problem statement of the sim-to-real gap in applying imitation learning to MPC. Section \uppercase\expandafter{\romannumeral3} details the proposed control framework. Section \uppercase\expandafter{\romannumeral4} presents the case study of the proposed framework using a cart-pole and vehicle collision avoidance systems. Conclusion and future work are provided in Section \uppercase\expandafter{\romannumeral5}.

\subsection{Notation}
Throughout the paper, the following set operations are used:  the Minkowski sum $A \oplus B := \{\, a+b \mid a \in A, \; b \in B \,\}$, the Pontryagin set difference $A \ominus B := \{\, x \mid x+b \in A, \; \forall b \in B \,\}$,  and the union of the sets $\bigcup_{i \in \mathcal{I}} A_i := \{\, x \mid x \in A_i, \, \forall i \in \mathcal{I} \,\}$.

\section{PROBLEM STATEMENT}

This section explains the process of applying imitation learning to MPC and where the sim-to-real gap occurs.

\subsection{Imitation learning of model predictive control}

The MPC's optimization problem calculates the optimal control input sequence $U^*(k)=[u^*_{0|k}, u^*_{1|k}, \cdots, u^*_{N-1|k}]^T$, minimizing the cost function with the constraint satisfaction over the prediction horizon as follows:
\begin{subequations}\label{eq_MPCopt}
\begin{align}
\underset{U(k)}{\min } \qquad &J(x(k), U(k))\\
\text { subject to } \quad \; \; &x_{i+1 | k}=f(x_{i|k}, u_{i|k},M) \label{nominal model1}\\
& x_{i|k} \in \mathbb{X}, u_{i|k} \in \mathbb{U}, x_{N|k} \in \mathbb{X}_f\\
& x_{0|k} = x(k)\\
& \text {for } i=0, \cdots, N-1,
\end{align}
\end{subequations}
\noindent where $J$ is the cost function; $f$ is the prediction model; $M$ denotes model parameters; $x$ and $u$ are the control state and input, respectively; $(\bullet)_{i|k}$ denotes the predicted value in step $k+i$ made in step $k$; $N$ is the prediction horizon; and $\mathbb{X}$, $\mathbb{U}$, and $\mathbb{X}_f$ are the constraint sets for the state and input. Then, the first element of the optimized input sequence $U^*(k)$ is used as the resultant control input $u_{\mathrm{MPC}}(x(k))$ at step $k$:
\begin{equation}
u_{\mathrm{MPC}}(x(k)) = u^*_{0|k}.
\label{eq_MPCinput}
\end{equation}

Applying imitation learning to MPC is to find the parameter $\theta$ of the DNN $\pi_\theta$ that best approximates the MPC by minimizing the expected error between $u_{\mathrm{MPC}}$ and $\pi_\theta$ in the target domain $\mathcal{T}$. By using the MPC's demonstrations in $\mathcal{T}$, $\theta$ is estimated as $\theta^*_{\mathcal{T}}$ through supervised learning with the widely used mean squared error loss function $\mathcal{L}$:
\begin{subequations}
\label{eq_IL_T}
\begin{gather}
\theta^*_{\mathcal{T}} = \underset{\theta}{\operatorname{argmin} \,} \mathbb{E}_{ 
 p(\xi|u_\mathrm{MPC},\mathcal{T})  } \left[ \mathcal{L}(\xi,\theta) \right], \\
\mathcal{L}(\xi,\theta) = \frac{1}{L_\mathrm{T}}\sum_{k=0}^{L_\mathrm{T}-1}\left| \left| u_{\mathrm{MPC}}(x(k)) - \pi_\theta (x(k)) \right| \right|^2_2,
\end{gather}
\end{subequations}
\noindent where $\xi=[x(0),x(1), \cdots, x(L_\mathrm{T}-1)]$ is the state trajectory with a length of $L_T$, driven by $u_\mathrm{MPC}$ in $\mathcal{T}$, and $p(\xi|u_\mathrm{MPC},\mathcal{T})$ is the probability distribution of $\xi$.

However, due to the difficulty of gathering MPC demonstrations in the target domain $\mathcal{T}$, a high-fidelity simulation, which is called the source domain $\mathcal{S}$, has been used to obtain them. Then, $\theta^*_{\mathcal{T}}$ is instead estimated as $\theta^*_{\mathcal{S}}$:
\begin{equation}
\theta^*_{\mathcal{S}} = \underset{\theta}{\operatorname{argmin} \,} \mathbb{E}_{ p(\xi|u_\mathrm{MPC},\mathcal{S})} \left[ \mathcal{L}(\xi,\theta) \right].
\label{eq_IL_S}
\end{equation}

\subsection{Sim-to-real gap}

This approach leads to the covariate shift problem due to the discrepancy between $\mathcal{T}$ and $\mathcal{S}$ that can be expressed as follows \cite{tagliabue2024efficient}:
\begin{equation}
\mathbb{E}_{ p(\xi|u_\mathrm{MPC},\mathcal{T})} \left[ \mathcal{L}(\xi,\theta^*_{\mathcal{S}}) \right] - \mathbb{E}_{ p(\xi|u_\mathrm{MPC},\mathcal{S})} \left[ \mathcal{L}(\xi,\theta^*_{\mathcal{S}}) \right].
\label{eq_shift}
\end{equation}

\noindent The above quantity indicates that an approximate DNN would not guarantee the designed performance in the target domain because it can encounter untrained states, a phenomenon commonly referred to as the sim-to-real gap problem. 

Therefore, attempts have been made to make the source domain encompass the target domain, such as applying random disturbances or sampling perturbations $d \sim p_\mathcal{T}(d)$ in the source domain, namely, DR. The approximate DNN $\theta^*_{\mathcal{S}.d}$ is then found as follows:
\begin{equation}
\theta^*_{\mathcal{S},d} = \underset{\theta}{\operatorname{argmin} \,} \mathbb{E}_{p_\mathcal{T}(d)} \left[ \mathbb{E}_{ p(\xi|u_\mathrm{MPC},\mathcal{S},d)} \left[ \mathcal{L}(\xi,\theta) \right] \right],
\label{eq_IL_E}
\end{equation}
\noindent with the probability distribution $p(\xi|u_\mathrm{MPC},\mathcal{S},d)$ denoting the transition probability driven by $u_\mathrm{MPC}$ in $\mathcal{S}$ under random perturbations $d$. DR has been widely used in control applications; however, determining the appropriate type and magnitude of perturbations remains challenging. Also, as the level of perturbation increases, the amount of required perturbed data grows significantly, increasing the complexity of collecting the training data. 

This study proposes a novel control framework that addresses the limitations of existing DRs, thereby significantly increasing data collection efficiency. Moreover, we propose an add-on governor in the framework that enables the overall controller to handle model parameter changes, ensuring that MPC constraints are satisfied under changing conditions and thereby allowing a less conservative MPC to be imitated.

\section{PROPOSED CONTROL FRAMEWORK}
This section briefly reviews the RTMPC, which inspired the proposed framework, and introduces the proposed control framework. Details on RTMPC can be referred to \cite{rawlings2017model}.

\subsection{Brief review of RTMPC}\label{Brief review of RTMPC}

RTMPC differs from the MPC in that it considers system dynamics disturbances $w(k)$:
\begin{equation}
\label{eq_disturbance}
x(k+1) =  f(x(k),u(k),M) + w(k),
\end{equation}
in the controller formulation. It results in a conservative control input that robustly satisfies the constraints in the presence of system disturbances, while the conventional MPC is vulnerable to constraint violations. 

\begin{figure}[t]
\centering
\includegraphics[width=0.95\columnwidth]{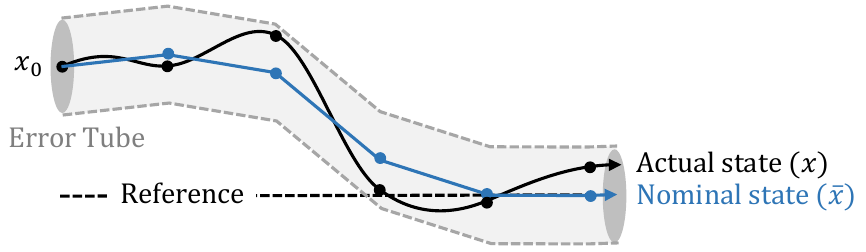}
\caption{RTMPC concept.}
\label{fig_RTMPC_structure}
\end{figure}

RTMPC's optimization problem computes a optimal control input sequence $\bar{U}^*(k)=[\bar{u}^*_{0|k}, \bar{u}^*_{1|k}, \cdots, \bar{u}^*_{N-1|k}]^T$, which minimizes the cost function while satisfying conservatively tightened constraints $\bar{\mathbb{X}}$, $\bar{\mathbb{U}}$, and $\bar{\mathbb{X}}_f$:
\begin{subequations}\label{eq_RTMPCopt}
\begin{align}
\underset{\bar{U}(k)}{\min } \qquad &J(\bar{x}(k), \bar{U}(k))\label{RTMPCopt1}\\
\text { subject to } \quad \; \; &\bar{x}_{i+1|k}=f(\bar{x}_{i|k}, \bar{u}_{i|k},M)\label{RTMPCopt2}\\
& \bar{x}_{i|k} \in \bar{\mathbb{X}}, \bar{u}_{i|k} \in \bar{\mathbb{U}}, \bar{x}_{N|k} \in \bar{\mathbb{X}}_f\label{RTMPCopt4}\\
& \bar{x}_{0|k} = \bar{x}(k)\label{RTMPCopt6}\\
& \text {for  } i=0, \cdots, N-1.\label{RTMPCopt7}
\end{align}
\end{subequations}
Then, the first element of the optimized input sequence is used as follows:
\begin{equation} \label{u_nom}
u_\mathrm{nom}(\bar{x}(k)) = \bar{u}^*_{0|k},
\end{equation}
which is called a nominal control input. The key difference of the optimization problems between MPC (\ref{eq_MPCopt}) and RTMPC (\ref{eq_RTMPCopt}) is that RTMPC's optimization problem is solved only with respect to the nominal state $\bar{x}$, as shown in (\ref{RTMPCopt6}). The nominal state is propagated using the prediction model (\ref{RTMPCopt2}) and the nominal input (\ref{u_nom}):
\begin{subequations}\label{nominal model}
\begin{gather}
\bar{x}(k+1) = f(\bar{x}(k), u_{\mathrm{nom}}(\bar{x}(k)),M) \label{nominal model_a}\\
\bar{x}(0) = x(0).
\end{gather}
\end{subequations}
\noindent which can be thought of as an open loop trajectory, made by the nominal input.

Only with the nominal input $u_\mathrm{nom}(\bar{x}(k))$, the actual state would propagate differently from the nominal state due to disturbances present in the system (\ref{eq_disturbance}). Therefore, RTMPC includes not only $u_\mathrm{nom}(\bar{x}(k))$ but also an ancillary controller $\kappa(x(k),\bar{x}(k))$ as follows:
\begin{equation}
u_\mathrm{RTMPC}(x(k)) = u_{\mathrm{MPC}}(\bar{x}(k)) + \kappa(x(k),\bar{x}(k)).
\label{eq_RTMPC actual input}
\end{equation}
The role of ancillary controller $\kappa(x(k),\bar{x}(k))$ is to guide the actual state to the nominal state by regulating the error between them, which is, stabilizing the following error dynamics:
\begin{equation}
\begin{aligned}
\label{eq_RTMPC error practical}
e(k+1) & = f(x(k), u_\mathrm{RTMPC}(x(k)),M) \\
& \quad  \,\,\,\,\, - f(\bar{x}(k), u_{\mathrm{nom}}(\bar{x}(k)),M) + w(k).
\end{aligned}
\end{equation}
As illustrated in Fig. \ref{fig_RTMPC_structure}, the stabilized error dynamics form a bounded region around the nominal state, referred to as the error tube. This tube represents the region within which the actual state remains while tracking the nominal trajectory. 

The nominal controller can ensure control stability for the nominal state by designing the cost function (\ref{RTMPCopt1}) as a Control Lyapunov Function (CLF). Then, by incorporating the concept of nominal and ancillary control inputs, RTMPC ensures the bounded stability of the actual state, maintaining it within a bounded error of the reference. Furthermore, by defining the terminal constraint set $\bar{\mathbb{X}}_f$ as a positive invariant set of the controlled nominal state and tightening the original constraints $\mathbb{X}$, $\mathbb{U}$ as tightened constraints $\bar{\mathbb{X}}$, $\bar{\mathbb{U}}$ by the size of the error tube, then RTMPC guarantees recursive feasibility, meaning that a stable constraint-satisfying control input always exists. 

In summary, RTMPC ensures both robust control stability and recursive feasibility in the presence of system disturbances, whereas conventional MPC cannot guarantee them. This is achieved by integrating the concepts of nominal and ancillary control inputs, along with a constraint-tightening strategy.

\subsection{Nominal and ancillary controllers}

The key concept of RTMPC that inspired this study is that the nominal controller operates solely on the nominal state $\bar{x}$, while the ancillary controller compensates for the error between the nominal state $\bar{x}$ and the actual state $x$. This suggests that if the DNN controller, approximating MPC, functions as the nominal controller, it only requires information from the nominal state distribution for training. Based on this idea, this study proposes the following definition and proposition.

\begin{definition}[ Nominal Model-based Domain $\mathcal{S_\mathrm{nom}}$]
\label{def_Nominal model based source domain}
A domain $\mathcal{S_\mathrm{nom}}$ in which the state $x$ is propagated by the nominal model, as in the following equation: 
\begin{equation}
x(k+1)=f(x(k),u(k),M),
\end{equation}
\end{definition}
\noindent is called the nominal model-based domain $\mathcal{S_\mathrm{nom}}$.

\begin{proposition}\label{Proposed imitation learning}
If the DNN $\pi_\theta$, approximating the MPC, is composed as a nominal controller in the overall control structure attached with an ancillary controller as follows:
\begin{equation}\label{Proposition control structure}
u(k) = \pi_\theta (\bar{x}(k)) + \kappa(x(k),\bar{x}(k)), 
\end{equation}
\noindent where $\bar{x}$ is the nominal state propagated by the nominal model, and $x$ is the actual state, then the target domain of the DNN controller becomes equal to the nominal model-based domain $\mathcal{S_\mathrm{nom}}$. 
\end{proposition}

\begin{proof}
The proof is trivial as the nominal DNN controller $\pi_\theta$ only receives the nominal state $\bar{x}$.
\end{proof}

That is, in the proposed framework, the DNN serves as the nominal controller and is responsible only for controlling the nominal state, while the ancillary controller ensures the actual state follows the nominal state. The key idea is that the DNN receives only the nominal state, making its target domain the same as the nominal model-based domain $\mathcal{S_\mathrm{nom}}$. Thus, training the DNN to approximate MPC requires only demonstrations from \(\mathcal{S_\mathrm{nom}}\), without accessing the full target domain \(\mathcal{T}\). This enables efficient imitation learning of MPC while mitigating the sim-to-real gap. Accordingly, using MPC demonstrations from \(\mathcal{S_\mathrm{nom}}\), the network parameters \(\theta\) are obtained as \(\theta^*_{\mathcal{S_\mathrm{nom}}}\) through supervised learning:
\begin{subequations}
\label{eq_IL_S_nom}
\begin{gather}
\theta^*_{\mathcal{S_\mathrm{nom}}} = \underset{\theta}{\operatorname{argmin} \,} \mathbb{E}_{ 
 p(\bar{\xi}|u_\mathrm{MPC},\mathcal{S_\mathrm{nom}})  } \left[ \mathcal{L}(\bar{\xi},\theta) \right], \\
\mathcal{L}(\bar{\xi},\theta) = \frac{1}{L_\mathrm{T}}\sum_{k=0}^{L_\mathrm{T}-1}\left| \left| u_{\mathrm{MPC}}(\bar{x}(k)) - \pi_\theta (\bar{x}(k)) \right| \right|^2_2,
\end{gather}
\end{subequations}
\noindent where $\bar{\xi}=[\bar{x}(0),\bar{x}(1), \cdots, \bar{x}(L_\mathrm{T}-1)]$ is the nominal state trajectory driven in $\mathcal{S_\mathrm{nom}}$ by $u_\mathrm{MPC}$, and $p(\bar{\xi}|u_\mathrm{MPC},\mathcal{S_\mathrm{nom}})$ is the probability distribution of $\bar{\xi}$ in $\mathcal{S_\mathrm{nom}}$.

\begin{figure}[t]
    \centering
    \subfloat[DR.]{%
        \includegraphics[width=0.44\columnwidth]{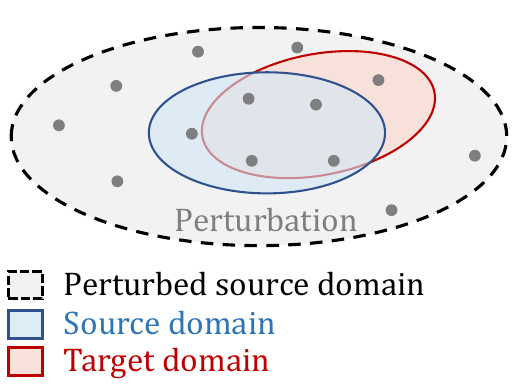} 
        \label{fig_DR}
    }
    \hfill 
    \subfloat[Proposed framework.]{%
        \includegraphics[width=0.44\columnwidth]{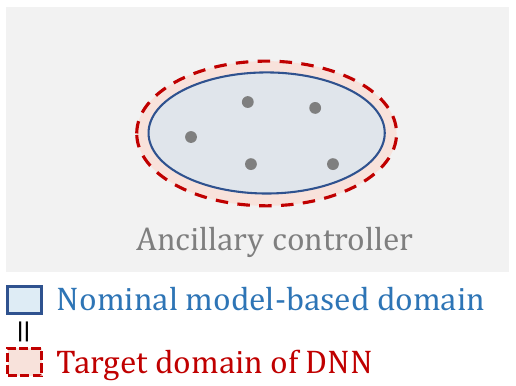} 
        \label{fig_Proposed}
    }
    \caption{Diagrams representing (a) Domain Randomization (DR) and (b) the proposed framework.}
    \label{fig_Diagram}
\end{figure}

Unlike DR, the DNN strictly operates within the source domain and is prevented from being exposed to untrained states. Therefore, additional perturbed sampling becomes unnecessary, significantly improving data collection efficiency. Note that defining the plant's operating range and sampling within the source domain are still required; however, the absence of external disturbance sampling distinguishes the proposed framework from DR-based methods.

Moreover, there is no need to create a high-fidelity source domain (simulation) since the nominal model used in $\mathcal{S_\mathrm{nom}}$ is already available from the MPC design process. Fig. \ref{fig_Diagram} graphically compares the proposed framework with conventional DR. While DR introduces perturbations in the source domain to encompass the target domain (Fig. \ref{fig_DR}), the proposed method redefines the DNN's target domain as the nominal model-based domain (Fig. \ref{fig_Proposed}).

It is noteworthy that the proposed approach does not eliminate the physical differences
between the source and target domains. Instead, these are compensated by the ancillary controller, allowing the DNN to operate solely within the source domain. In other words, the sim-to-real gap is no longer a "problem directly addressed by the DNN", but a "problem addressed by the ancillary controller." To be more specific, the sim-to-real gap is represented as the state propagation difference between the actual and nominal states $e(k)=x(k)-\bar{x}(k)$, having the error dynamics as follows:
\begin{equation}
\begin{aligned}
\label{eq_proposed error practical}
e(k+1) & = f(x(k), u(k),M) \\
& \quad  \,\,\,\,\, - f(\bar{x}(k), \pi_{\theta}(\bar{x}(k)),M) + w(k),
\end{aligned}
\end{equation}
which is a similar form to (\ref{eq_RTMPC error practical}).

\subsubsection{Remark}
As discussed in Section~\ref{Brief review of RTMPC}, RTMPC ensures robust stability when the nominal controller stabilizes the nominal state, and a stable ancillary controller is applied. The proposed framework follows the same principle: if the imitated MPC and the ancillary controller are both stable, the overall closed-loop system remains robustly stable. This stability depends on the DNN's ability to accurately approximate the MPC with small approximation errors, which is a difficult challenge in the learning-based area \cite{alsmeier2024imitation,yang2022guaranteed}. However, because the proposed framework narrows the DNN's operating domain to the source domain, it enhances the reliability of this approximation and strengthens the overall stability.

Inspired by existing RTMPC studies\cite{lee2023robust,jeong2023tube}, an effective way to design a stable ancillary controller is to set it as a full-state feedback controller:
\begin{equation}\label{}
\kappa(x(k),\bar{x}(k)) = K(x(k) - \bar{x}(k)),
\end{equation}
\noindent where $K$ is set to make $A(k)+B(k)K$ Hurwitz-stable for all the possible linearized system matrices $A(k) = \nabla_xf|_{\bar{x}(k),\bar{u}(k),M}$ and $B(k) = \nabla_uf|_{\bar{x}(k),\bar{u}(k),M}$, derived by the following linearized error dynamics:
\begin{equation}
\label{eq_RTMPC lin error practical}
\begin{aligned}
e(k+1) = (A(k)+ B(k)K)&e(k)\\
& +  O(\left| \left|e(k) \right| \right|^2) + w(k),
\end{aligned}
\end{equation}
\noindent where $O(\left| \left|e(k) \right| \right|^2)$ denotes the linearization error. Even if the system is underactuated or the states are partially measurable, an observer-based full-state feedback form of the ancillary controller can still be designed, provided that the system is controllable and observable. 

Note that the ancillary controller is not limited to a full-state feedback form. Other controller forms can also be employed to regulate the error between the actual and nominal states. For instance, a nonlinear controller such as a FeedBack Linearization Controller (FBLC) can be applied for systems exhibiting significant nonlinearities. Robust controllers, such as an H-infinity controller or a Sliding Mode Controller (SMC), can be used for systems with significant uncertainties.

\subsection{Input refinement governor} \label{Sec input refinement governor}

As discussed above, the sim-to-real gap still exists, and to ensure that the actual states satisfy the original constraints, it is necessary to apply conservative constraint tightening in the imitated MPC. A major drawback of imitation learning is that once the DNN approximates the MPC, it becomes difficult to adapt to variations in the plant's model parameters. Hence, many studies have treated parameter variations as external disturbances, leading to overly conservative control behavior.  
However, various parameter estimation methods\cite{hong2014novel,choi2013linearized, han2018estimation} have been proposed, and treating such known variations merely as disturbances is inefficient.   

Therefore, the proposed framework introduces an add-on component, referred to as the \textit{input refinement governor}, which refines the control input from $u$ to $u^*$ actively utilizing the estimated model parameters instead of the DNN controller. This approach enables a less conservative constraint design and ultimately enhances the overall control performance. In this subsection, the tightening set \(\mathbb{Y}_{\mathbb{S}_\infty}\) is mathematically derived to represent the required degree of constraint tightening, and it is shown that this set is reduced when the proposed governor is applied.

The original constraint sets $\mathbb{X}$ and $\mathbb{U}$ can be integrated to $\mathbb{Y}$, defined by the following function $g(\bullet)$:
\begin{equation}
\mathbb{Y} = \mathbb{X} \times \mathbb{U}:= \{(x, u) \mid g(x, u, \tilde{M})\leq 0 \},
\label{eq_Constraint inequility}
\end{equation}
\noindent where $\tilde{M}$ represents the plant's changed model parameters. Then, the tightened constraint set can be defined using the tightening parameter $\gamma \geq 0$, as follows:
\begin{equation}
\bar{\mathbb{Y}} = \bar{\mathbb{X}} \times \bar{\mathbb{U}}:= \{(\bar{x}, \bar{u}) \mid g(\bar{x}, \bar{u}, M)\leq -\gamma \}.
\label{eq_Tightened constraint inequility}
\end{equation}
\noindent Note that the approximated MPC is designed to satisfy the constraints made for the nominal model parameter $M$ as in the above equation. 

Tightening parameter $\gamma$ is obtained by subtracting the tightening set $\mathbb{Y}_{\mathbb{S}_\infty}$ from the original constraint set $\bar{\mathbb{Y}}$, using Pontryagin set difference $\ominus$ \cite{lee2023robust}:
\begin{equation}
\bar{\mathbb{Y}} = \mathbb{Y} \ominus \mathbb{Y}_{\mathbb{S}_\infty}.
\end{equation} 
\noindent Referring to \cite{rawlings2017model}, the tightening set $\mathbb{Y}_{\mathbb{S}_\infty}$ is defined by the upper bound, specifically the convex hull, of the difference between the left-hand sides of the inequalities in (\ref{eq_Constraint inequility}) and (\ref{eq_Tightened constraint inequility}):
\begin{equation}
\label{eq_RTMPC Y error}
e_g(k) = g(x(k),u(k),\tilde{M}) - g(\bar{x}(k),u_\mathrm{MPC}(\bar{x}(k)),M).\\
\end{equation}
\noindent This implies that the constraints should be conservatively tightened according to the resulting difference in $g(\bullet)$. Various methods of deriving $\mathbb{Y}_{\mathbb{S}_\infty}$ by (\ref{eq_RTMPC Y error}) have been derived in the RTMPC field. In the case of linear systems, the tightening set can be represented as a polytope, which is easily obtained using set operations \cite{mayne2005robust}, and for nonlinear systems, the sublevel set of Lyapunov functions is defined that upper-bounds the tightening set\cite{mayne2011tube}.

In this study, to intuitively demonstrate that the proposed governor reduces the tightening set, (\ref{eq_RTMPC Y error}) is linearized with respect to the nominal state trajectory, and $\mathbb{Y}_{\mathbb{S}_\infty}$ is derived using set operations as in\cite{mayne2005robust}. Although this linearization approach gives only a rough derivation, it provides a reasonable approximation that reveals the effectiveness of the proposed governor. The resulting linearization error is proportional to the deviation between the nominal and actual states. Since the ancillary controller regulates this deviation, the linearization error is roughly assumed to be bounded and negligibly small in the derivations.

By applying (\ref{Proposition control structure}) into (\ref{eq_RTMPC Y error}), the following linearized form is obtained:
\begin{equation}
\begin{aligned}
e_g(k) \approx C(k)e(k)+D(k)&\kappa(x(k),\bar{x}(k)) \\ 
& + w_M(k)  + w_{\pi_\theta}(k),
\end{aligned}
\label{eg linear}
\end{equation}
where \(e(k) = x(k) - \bar{x}(k)\), and the linearized matrices are defined as 
\(C(k) = \nabla_x g|_{\bar{x}(k),\pi_\theta(\bar{x}(k)),M}\) and 
\(D(k) = \nabla_u g|_{\bar{x}(k),\pi_\theta(\bar{x}(k)),M}\).  
The term \(w_M(k)\) represents the effect of model parameter variations:
\begin{equation}
w_M(k) = \nabla_M g|_{\bar{x}(k),\pi_\theta(\bar{x}(k)),M} (\tilde{M} - M),
\end{equation}
while \(w_{\pi_\theta}(k)\) represents the effect of DNN approximation errors:
\begin{equation}
w_{\pi_\theta}(k) = \nabla_u g|_{\bar{x}(k),\pi_\theta(\bar{x}(k)),M}
(\pi_\theta(\bar{x}(k)) - u_\mathrm{MPC}(\bar{x}(k))).
\end{equation}
The approximation symbol (\(\approx\)) in (\ref{eg linear}) arises from linearization errors, which are assumed to be negligible in this study.  

Quantifying or upper-bounding the DNN approximation error \(w_{\pi_\theta}(k)\) remains a challenging task. Several studies have proposed algorithms that estimate such bounds based on the Lipschitz properties of neural networks \cite{alsmeier2024imitation,yang2022guaranteed}. Rather than directly computing this bound, this study focuses on describing how \(w_{\pi_\theta}(k)\) and other elements of the tightening set are affected by the proposed governor, and demonstrates how the governor effectively reduces the overall tightening set in the framework.

By accumulating $e_g(k)$ over the infinite time horizon starting from step $n \in \mathcal{Z}_{\geq 0}$ (non-negative integer), the tightening set $\mathbb{Y}_{\mathbb{S}_\infty}$ is obtained as follows using the Minkowski sum $\oplus$ \cite{lee2023robust}:
\begin{equation}
\label{eq_UnFilter RTMPC Y invariant set}
\mathbb{Y}_{\mathbb{S}_\infty} \approx  \bigcup_{n \in \mathcal{Z}_{\geq 0}}  \bigoplus^{\infty}_{k=n} ( C(k) \, \mathbb{S}_{n,k} \oplus D(k)\mathbb{K})\oplus\mathbb{W}_{M}\oplus\mathbb{W}_{\pi_{\theta}},
\end{equation}
where $\mathbb{W}_{M}$ and $\mathbb{W}_{\pi_\theta}$ are the set for $w_M(k)$ and $w_{\pi_\theta}(k)$, respectively, $\mathbb{K}$ is the set for the ancillary controller $\kappa(x(k),\bar{x}(k))$ and $\mathbb{S}_{n,k}$ is the error set for $e(k)$ from step $n$ to $k$. The error set $\mathbb{S}_{n,k}$ can be derived by the error dynamics (\ref{eq_proposed error practical}), which can be re-arranged as follows:
\begin{equation}
e(k+1) = \Phi(k)e(k) + w(k), 
\end{equation}
using a time-varying system matrix $\Phi(k)$. Therefore, $\mathbb{S}_{n,k}$ can be expressed as follows:
\begin{equation}
\label{error_set}
\mathbb{S}_{n,k} =  \bigoplus^{k}_{n} \Phi(n)^{k-n} \, \mathbb{W},
\end{equation}
where $\mathbb{W}$ represents the disturbance set for $w(k)$.

According to (\ref{eq_UnFilter RTMPC Y invariant set}), the tightening set $\mathbb{Y}_{\mathbb{S}_\infty}$ consists of the sets $\mathbb{S}_{n,k}$, $\mathbb{K}$, $\mathbb{W}_{M}$, and $\mathbb{W}_{\pi_\theta}$. Since $\mathbb{W}_{\pi_\theta}$ arises due to the DNN approximation error, the objective of the proposed governor is to reduce the sizes of the remaining sets $\mathbb{S}_{n,k}$, $\mathbb{K}$, and particularly $\mathbb{W}_{M}$ which reflects model parameter variations and tends to make the constraints overly conservative.

\begin{definition}[ Input refinement governor]\label{Def_governor}
The input refinement governor computes a refined input $u^*(k)$ that satisfies
\begin{equation}\label{eq_Model-based filter}
    g(x(k),u^*(k),\tilde{M}) = g(\bar{x}(k),u(k),M).
\end{equation}
\end{definition}

The proposed governor can be intepreted as trying to make the constraint satisfied for the changed model parameters, which is making $g(\bullet,\tilde{M})\leq0$ for the changed parameters $\tilde{M}$ by matching $g(\bullet,\tilde{M})=g(\bullet,M)$. The governor is assumed to be tractable, meaning that the refined control input, obtained as the solution of (\ref{eq_Model-based filter}), can be computed within a reasonable time. 

Followed by the condition for the refined input that it reduces the new ancillary controller set $\mathbb{K}^*$ when applying $u^*(k)$ instead of $u(k)$, meaning:
\begin{equation} \label{Assumption_eq}
\mathbb{K}^* \subseteq \mathbb{K},
\end{equation}
it is shown that the required tightening set is reduced when applying the proposed governor by the following proposition.

\begin{proposition}\label{Proposed filter reduce tightning set}
Under the control structure of Proposition \ref{Proposed imitation learning}, if the control input $u(k)$ is refined as $u^*(k)$ by the governor (\ref{eq_Model-based filter}) that satisfies condition (\ref{Assumption_eq}), then the tightening set $\mathbb{Y}^*_{\mathbb{S}_\infty}$ resulting from the refined control input $u^*(k)$ is the subset of the tightening set $\mathbb{Y}_{\mathbb{S}_\infty}$ from the control input $u(k)$:
\begin{equation}
\mathbb{Y}^*_{\mathbb{S}_\infty} \subseteq \mathbb{Y}_{\mathbb{S}_\infty}.
\end{equation}
\end{proposition}

\begin{proof}
Since the refined control input $u^*(k)$ is applied, the tightening set $\mathbb{Y}^*_{\mathbb{S}_\infty}$ can be redefined as the upper bound of the following difference:
\begin{equation}
\label{eq_RTMPC Y error_Pro}
e^*_g(k) = g(x(k),u^*(k),\tilde{M}) - g(\bar{x}(k),\bar{u}(k),M),
\end{equation}
which can be rearranged as follows by applying (\ref{eq_Model-based filter}):
\begin{equation}
\label{eq_Filter RTMPC Y error}
e^*_g(k) = g(\bar{x}(k),u(k),M) - g(\bar{x}(k),\bar{u}(k),M).
\end{equation}
Therefore, the above equation can be linearized as:
\begin{equation}
e^*_g(k)\approx D(k)\kappa(x(k),\bar{x}(k)) + w_{\pi_\theta}(k).
\end{equation}
Then, the tightening set $\mathbb{Y}^*_{\mathbb{S}_\infty}$ resulting from the refined control input $u^*(k)$ can be derived as:
\begin{equation}
\label{eq_Filter RTMPC Y invariant set}
\mathbb{Y}^*_{\mathbb{S}_\infty} \approx \bigcup_{n\in\mathcal{Z}_{\geq 0}} \bigoplus^{\infty}_{k=n} D(k)\mathbb{K}^* \oplus \mathbb{W}_{\pi_{\theta}},
\end{equation}
\noindent where  $\mathbb{K}^*$ is the new set for the ancillary controller $\kappa(x(k),\bar{x}(k))$ when refined input $u^*$ is applied. To show that the tightening set $\mathbb{Y}_{\mathbb{S}_\infty}$ is subset  of the tightening set $\mathbb{Y}^*_{\mathbb{S}_\infty}$, define $\tilde{\mathbb{Y}}_{\mathbb{S}_\infty}$ as an intermediate tightening set as follows:
\begin{equation}
\tilde{\mathbb{Y}}_{\mathbb{S}_\infty}  =  \bigcup_{n\in\mathcal{Z}_{\geq 0}} \bigoplus^{\infty}_{k=n} D(k)\mathbb{K}\oplus\mathbb{W}_{\pi_{\theta}}.
\end{equation}
The following inclusion holds because $\mathbb{K}^* \subseteq \mathbb{K}$:
\begin{equation}\label{proof1}
\mathbb{Y}^*_{\mathbb{S}_\infty} \subseteq \tilde{\mathbb{Y}}_{\mathbb{S}_\infty}.
\end{equation}
Also, because the terms $\mathbb{S}_{n,k}$ and $\mathbb{W}_{M}$ are included in $\mathbb{Y}_{\mathbb{S}_\infty}$ compared to $\tilde{\mathbb{Y}}_{\mathbb{S}_\infty}$, the following inclusion holds:
\begin{equation}\label{proof2}
\tilde{\mathbb{Y}}_{\mathbb{S}_\infty} \subseteq \mathbb{Y}_{\mathbb{S}_\infty}.
\end{equation}

By (\ref{proof1}) and (\ref{proof2}), it is confirmed that $\mathbb{Y}^*_{\mathbb{S}_\infty}$ is a subset of $\mathbb{Y}_{\mathbb{S}_\infty}$, meaning that the required tightening set $\mathbb{Y}^*_{\mathbb{S}_\infty}$ is reduced from $\mathbb{Y}_{\mathbb{S}_\infty}$.

\end{proof}

In summary, the proposed governor refines the control input based on the changed model parameters, instead of the DNN controller. This enables a less conservative MPC to be approximated by the DNN while ensuring that MPC constraints remain robustly satisfied even under variations in model parameters.

Practically, (\ref{Assumption_eq}) can also be thought of as the following condition:
\begin{equation} \label{eq_disturbance new}
\mathbb{W}^* \subseteq \mathbb{W},
\end{equation}
\noindent where $\mathbb{W^*}$ represents the new disturbance set for the discrepancy $w^*(k)$ between the plant and the nominal model when applying the refined input $u^*(k)$ to the plant:
\begin{equation}
\label{eq_new disturbance}
x(k+1) =  f(x(k),u(k), M) + w^*(k).
\end{equation}

(\ref{eq_disturbance new}) means that the refined input makes the behavior of the plant closer to that of the nominal model, which naturally reduces the magnitude of the ancillary controller ($\mathbb{K}^* \subseteq \mathbb{K}$). (\ref{eq_disturbance new}) can be used instead of (\ref{Assumption_eq}) when designing the governor, as it is easier to verify the condition before designing the ancillary controller by comparing state propagations between the plant and the nominal model. 

\begin{figure}[t]
\centering
\includegraphics[width=1\columnwidth]{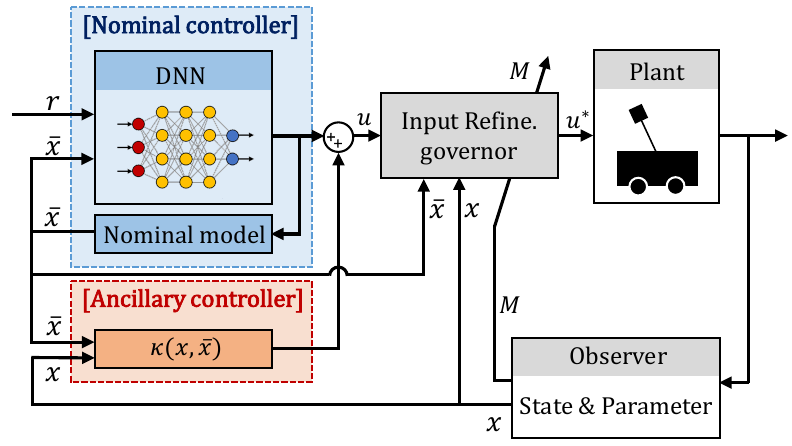}
\caption{The overall structure of the proposed control framework.}
\label{fig_Control_structure}
\end{figure}

\subsubsection{Remark}

Fig. \ref{fig_Control_structure} illustrates the overall structure of the proposed control framework. The DNN controller is composed as a nominal controller, operating within the same environment as the source domain $\mathcal{S_\mathrm{nom}}$. An ancillary controller ensures that the actual plant follows the nominal model, while a governor refines the control input to handle model parameter variations. 

Notably, the overall closed-loop system stability is maintained under condition (\ref{Assumption_eq}), even with the input refinement governor. As mentioned earlier, stability is guaranteed when both the imitated MPC and the ancillary controller are stable. Moreover, as the governor reduces the magnitude of the feedback control input (\ref{Assumption_eq}), the actual state follows the nominal state more closely, thereby reinforcing overall closed-loop stability. 

The proposed framework assumes the existence of a governor that satisfies the form (\ref{eq_Model-based filter}) and conditions (\ref{Assumption_eq}) or (\ref{eq_disturbance new}). Due to the nonlinearity of $g(\bullet)$ or state estimation errors (measurement noise), only an approximate solution may be obtained in practice. Nevertheless, the governor still computes a refined input that suppresses variations in $g(\bullet)$ caused by parameter changes, thereby retaining its effectiveness. To provide insight into its practical implementation, this study presents two case studies that demonstrate how the proposed framework can be generalized to other control problems.

\section{CASE STUDY}
The proposed control framework was validated in two case studies: cart-pole control and vehicle collision avoidance control. Each analyzes the principles of the proposed framework in detail and shows the results of applying it to a practical vehicle control case. 

\subsection{Cart-pole system}

\begin{figure}[t]
    \centering
    \includegraphics[width=0.3\textwidth]{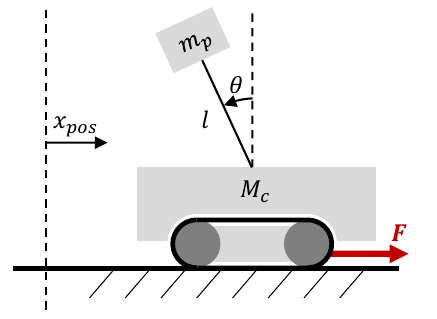}
    \caption{Cart-pole system.}
    \label{fig_Ch3_Veri_Cartpole}
    \vspace{0mm}
\end{figure}

For the cart-pole system, shown in Fig. \ref{fig_Ch3_Veri_Cartpole}, the control objective is to regulate both cart position and pole angle to the origin while ensuring the wheel force does not exceed the friction limit. To achieve this, an MPC was designed, followed by imitation learning, and applied to the proposed control framework.

The system dynamics for the cart-pole is given by:
\begin{subequations}
\label{eq_cartpole equation of motion}
\begin{align}
\Ddot{x}_\mathrm{pos} = \frac{F-(m_p l \dot{\theta}^2 -m_p g \cos \theta) \sin \theta}{M_c+m_p \sin ^2 \theta},\quad\,\,\,& \\
\Ddot{\theta}=\frac{\cos \theta ( F-(m_p l \dot{\theta}^2-\left(M_c+m_p\right) g )\sin \theta ) }{l(M_c+m_p \sin ^2 \theta)},&
\end{align}
\end{subequations}
where $x_\mathrm{pos}$ is the cart position, $\theta$ is the pole angle, $M_c$ and $m_p$ are cart and pole masses, $l$ is the pole length, and $g$ is the gravitational acceleration. $F$ is the control input representing the force applied by the cart's wheel. The system can be represented by the state $x=[x_\mathrm{pos}, \dot{x}_\mathrm{pos},\theta,\dot{\theta}]^T$, control input $u=F$, and model parameters ${M} = [{M}_c, {m}_p]$. 

The control objective is to regulate both $x_\mathrm{pos}$ and $\theta$ to the origin from $x(0)=[3,0,0,0]^T$ while ensuring the wheel force does not exceed the friction limit. To formulate the MPC's optimization problem, the prediction model (nominal model) was made as follows:
\begin{equation}
x(k+1)  =  f(x(k),u(k), M),
\end{equation}
which was derived from (\ref{eq_cartpole equation of motion}) and discretized using the 4th-order Runge-Kutta method (RK4) with sampling time 0.01s. The nominal model parameter was set as $M=[4\mathrm{kg, }1\mathrm{kg}]$. A cost function penalizing the states and excessive control inputs was composed as:
\begin{equation}
J(x(k), U(k))=\sum_{i=0}^{N-1} (x_{i|k}^T Q x_{i|k}+u_{i|k}^T R u_{i|k}),
\end{equation}
with the horizon of $N=50$ and $Q=\mathrm{diag}([20,0,5,0])$, $R=0.001$. Finally, constraints that prevent the wheel force from exceeding the friction limit were formulated as follows:
\begin{equation}\label{eq_friction limit}
\bar{\mathbb{Y}} = \bar{\mathbb{X}} \times \bar{\mathbb{U}}= \{(\bar{x}, \bar{u}) \mid \left| \frac{F}{\mu F_z}\right| - 1 \leq -\gamma \} ,
\end{equation}
where the normal force $F_z$ is given by:
\begin{equation}
F_z(x,u,M) = M_c g + m_p(g - l(\Ddot{\theta}\sin\theta + \dot{\theta}^2 \cos\theta)),
\label{eq_cartpole Fz}
\end{equation}
with road friction $\mu=0.5$. The tightening parameter was set as $\gamma = 0.2$, which is derived in detail later. Finally, the designed MPC, with a sufficiently long prediction horizon, ensured nominal stability and successfully regulated the system states to the origin for the nominal model.

\subsubsection{Proposed control framework}

By imitation learning, the designed MPC was approximated to a fully connected DNN using supervised learning, with 4 hidden layers of 17 nodes each:
\begin{equation}
u_\mathrm{MPC}(\bullet) \,\rightarrow \, \pi_{\theta^*_{\mathcal{S_\mathrm{nom}}}}(\bullet),
\end{equation}
and implemented into the proposed framework. Because the goal of this case study is to verify the proposed framework's great data collection efficiency of handling the sim-to-real gap, only a single MPC-demonstrated trajectory $\bar{\xi}$ was collected from $\mathcal{S_\mathrm{nom}}$, starting from the initial state $x(0)=[3,0,0,0]^T$ to the origin for 100 steps. 

The ancillary controller was designed as a full-state feedback controller, with the typical Linear Quadratic Regulator (LQR) gain $K=[5.7,10.3,-139.9, -41.7]$ obtained by solving the Discrete Algebraic Riccati Equation (DARE) for the linearized model equation at the origin. Under the proposed framework, the control input $u(k)$ was formulated as:
\begin{equation}
u(k) = \pi_{\theta^*_{\mathcal{S_\mathrm{nom}}}}(\bar{x}(k)) + K(x(k) - \bar{x}(k)).
\label{eq_cartpole actual input}
\end{equation}

Finally, the input refinement governor (\ref{eq_Model-based filter}) leads to the following form regarding the friction limit constraints (\ref{eq_friction limit}):
\begin{equation}\label{eq_cartpole filter1}
g_\mathrm{Fr.}(x(k),u^*(k),\tilde{M}) = g_\mathrm{Fr.}(\bar{x}(k),u(k),M),
\end{equation}
where $g_\mathrm{Fr.}(\bullet)$ is defined as follows:
\begin{equation}
g_\mathrm{Fr.}(x(k),u(k),M) =\left| \frac{u(k)}{\mu F_z(x(k),u(k),M)} \right| - 1.
\end{equation}
Then, the refined input $u^*$ satisfying (\ref{eq_cartpole filter1}) can be calculated by the following equation:
\begin{equation}
\label{eq_cartpole filter2}
u^*(k) = u(k) \times \frac{ F_z(x(k),u^*(k),\tilde{M})}{F_z(\bar{x}(k),u(k),M)},
\end{equation}
which can be physically interpreted as refining the wheel force proportionally to changes in the normal force; that is, if the cart's mass increases, the force input is also increased accordingly. Intuitively, it is expected to satisfy (\ref{eq_disturbance new}) because it refines the control input to match the changed mass, as shown in the later graphs.

The Cart-pole was controlled in the target domain, where a disturbance $d$ within the range $[-5,5]$ was randomly applied to the control input ($F=u+d$), creating a gap to the source domain. The disturbance represents parametric uncertainties and external friction effects that are not considered in the source domain. The actual model parameter $\tilde{M}$ of the cart-pole was first set to be the same as the nominal ones $M$, and the case under model parameter variations is introduced later.

DR baselines were chosen for comparison with the proposed control framework, including the conventional DR method \cite{tobin2017domain} and the recently proposed tube-based DR \cite{tagliabue2024efficient}. While other transfer learning baselines, such as domain adaptation and meta-learning, have also been explored to address the sim-to-real gap, they generally require access to the target domain for adaptation or policy fine-tuning. In contrast, the proposed framework compensates for the sim-to-real gap without requiring access to the target domain, relying solely on the source domain. Therefore, conceptually similar DR-based approaches were selected for comparison.

Both DR baselines approximated the RTMPC controller:
\begin{equation}
u_{\mathrm{RTMPC}}(x(k)) = u_{\mathrm{MPC}}(\bar{x}(k)) + K(x(k)-\bar{x}(k)),
\end{equation}
where $u_\mathrm{MPC}(\bar{x}(k))$ and $K$ were designed as same as in the proposed framework. This eventually results in the same control inputs as the proposed framework, intended to make the baselines match its control performance. DR baselines collected additional MPC demonstrations around the trajectory $\bar{\xi}$, where sparse sampling was used for DR w/ tube, and perturbed 10 trajectories resulted by applying random disturbances within the range of $[-5,5]$ were collected for the conventional DR, denoted as DR (10 Traj.). In these baselines, the DNN alone controlled the cart-pole, also handling the sim-to-real gap.

\subsubsection{Results: sim-to-real gap}

All MPC-approximated controllers achieved a significant reduction in computation time per step, from 0.5s for the original RTMPC to approximately 0.6ms. This demonstrates a major advantage of the MPC-based imitation learning framework, which can substantially improve computational efficiency while maintaining the control performance of the original controller.

\begin{figure}[t!]
    \centering
    \subfloat[Controlled cart position compared with DR baselines.]{%
        \includegraphics[width=1\columnwidth]{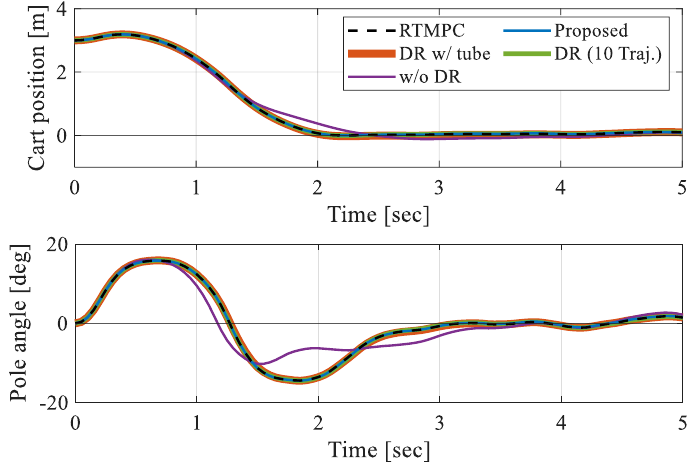} 
        \label{fig_Posi_data}
    }
    \vfill 
    \subfloat[DNN and ancillary control inputs of the proposed framework (blue) and DNN control inputs of the baselines (red, green) in the target domain.]{%
        \includegraphics[width=1\columnwidth]{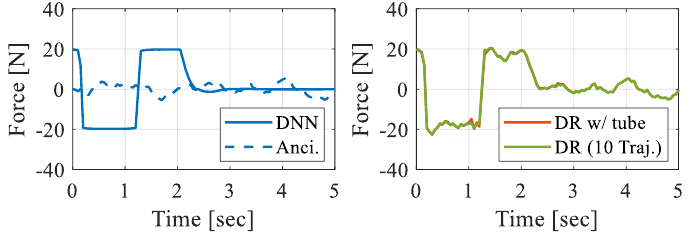} 
        \label{fig_Input_data}
    }
    \caption{Controlled cart position and the control inputs in the target domain.}
    \label{fig__data}
\end{figure}

\begin{figure}[t!]
    \centering
    \includegraphics[width=1\columnwidth]{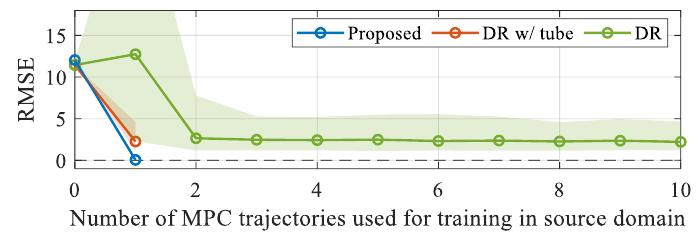}
    \caption{Root Mean Squared Error (RMSE) between the inputs of the original controller and the approximated DNN in the target domain. For the case of a zero MPC trajectory, $\pi_\theta=0$ is applied for calculating the RMSE value.}
    \label{fig_RMSE_data}
    \vspace{0mm}
\end{figure}

Fig. \ref{fig__data} illustrates the key concept of the proposed framework by showing the controlled cart-pole's position and pole angle in the target domain (Fig. \ref{fig_Posi_data}) and the control inputs generated by the proposed framework and baselines (Fig. \ref{fig_Input_data}). As shown in Fig. \ref{fig_Posi_data}, both the proposed method and the DR baselines successfully regulated the cart-pole to the origin, showing results that are almost aligned with those of the original RTMPC, thereby overcoming the sim-to-real gap. This is in contrast to the case when DR is not used (w/o DR), where an oscillating trajectory resulted from the gap. As shown in Fig. \ref{fig_Input_data}, noisy control inputs were found in the baselines since the DNN handled all disturbances in the target domain. In contrast, the DNN in the proposed framework generated cleaner control inputs, as it controlled only the nominal state. 

\begin{table}[t!]
\caption{Comparison of Root Mean Square Error (RMSE).}
\label{tab_corr rms}
\begin{center}
\begin{tabular}{|c|c|c|c|}
\hline & Proposed & DR w/ tube & DR (10 Traj.) \\
\hline 
RMSE & 0.0525 & 2.263 & 2.298 \\
\hline
\end{tabular}
\end{center}
\end{table}

To evaluate performance in addressing the sim-to-real gap, the Root Mean Squared Error (RMSE) between the DNN's control inputs and the original controller's control inputs that would have resulted in the target domain was analyzed. A smaller value indicates that the DNN represented the original controller better beyond the sim-to-real gap. Fig. \ref{fig_RMSE_data} shows the RMSE value plotted by the amount of collected data. The results of 100 operations with different random disturbances are represented as the shaded area, with the average value plotted as the solid line.

The proposed framework achieved the best sim-to-real transfer without any access to the target domain or extensive data augmentation. As shown in Fig.~\ref{fig_RMSE_data}, the framework achieved nearly zero RMSE even with a single MPC demonstration, because the DNN operates strictly within the source domain. This highlights a key distinction from DR methods, which rely on random sampling of disturbances, as well as domain adaptation and meta-learning methods, that require access to the target domain for pre-alignment or further adaptation. DR baselines did not match the proposed method's performance, even with more samples, because the sampled disturbances did not fully capture the true disturbances in the target domain. The tube-based DR showed better data efficiency than conventional DR, achieving greater performance with a single trajectory but still exhibiting steady errors for the same reason. Overall, the results summarized in Table~\ref{tab_corr rms} confirm that the proposed structure effectively bridges the sim-to-real gap while maintaining superior data efficiency compared to all baseline methods.

\subsubsection{Results: model parameter changes}

To evaluate the proposed framework's capability to handle variations in model parameters, additional experiments were conducted in the target domain with changed parameters, assuming they were estimated by existing parameter estimation methods. The changed model parameters were set to $\tilde{M}=[6\mathrm{kg, }0.5\mathrm{kg}]$, differing from the nominal model parameters $M=[4\mathrm{kg, }1\mathrm{kg}]$.

\begin{figure}[t!]
    \centering
    \includegraphics[width=1\columnwidth]{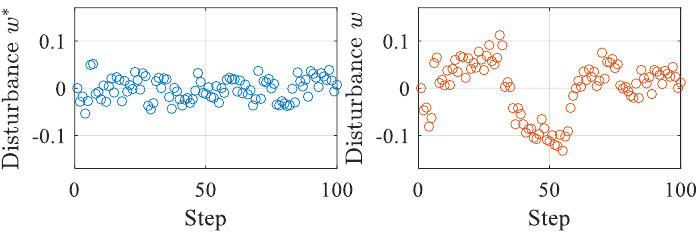} 
    \caption{System disturbance of $\dot{x}_\mathrm{pos}$ with (blue) and without (red) the governor.}
    \label{fig_Dist_govern}
    \vspace{0mm}
\end{figure}

First, it was verified that the designed governor reduced the tightening set, satisfying the condition (\ref{eq_disturbance new}). Fig. \ref{fig_Dist_govern} shows one of the state disturbances (for $\dot{x}_\mathrm{pos}$) encountered in the target domain, during control with and without the governor. As shown in the figure, the refined control input reduced system disturbances, indicating that it brought the plant's behavior closer to that of the nominal model. Therefore, by Proposition \ref{Proposed filter reduce tightning set}, the proposed governor reduces the tightening set.

\begin{figure}[t!]
    \centering
    \includegraphics[width=1\columnwidth]{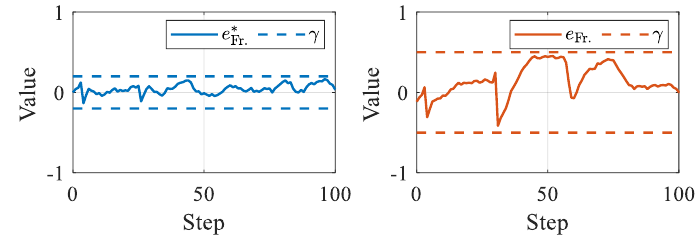} 
    \caption{Required tightening parameter $\gamma$ with (blue) and without (red) the governor.}
    \label{fig_Set_govern}
    \vspace{0mm}
\end{figure}

\begin{figure}[t!]
    \centering
    \subfloat[Resulted control inputs with (blue) and without (red) the governor.]{%
        \includegraphics[width=1\columnwidth]{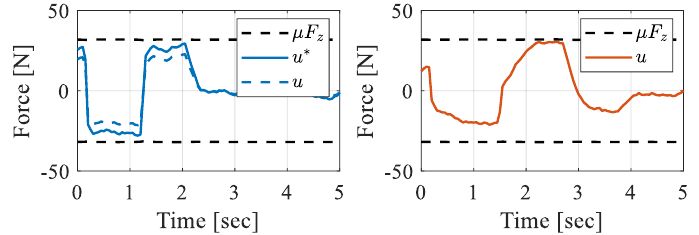} 
        \label{fig_Input_govern}
    }
    \vfill 
    \subfloat[Controlled cart position with and without the governor.]{%
        \includegraphics[width=1\columnwidth]{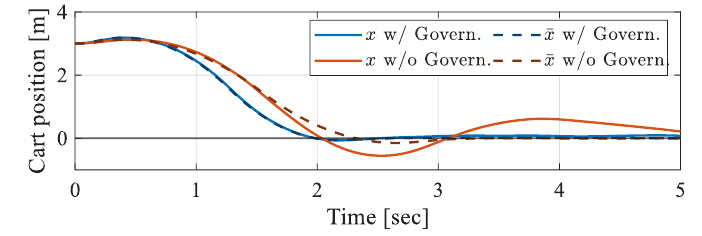} 
        \label{fig_Posi_govern}
    }
    \caption{The results of controlled cart position and control inputs with (blue) and without (red) the governor.}
    \label{fig__govern2}
\end{figure}

The resulting reduction in the tightening parameter was also checked. Referring to the meaning of the tightening set in Section \ref{Sec input refinement governor}, the tightening parameter $\gamma$ must be at least as big as the absolute value of $e_g$ in (\ref{eq_RTMPC Y error}). Therefore, series of $e_\mathrm{Fr.}$ and $e^*_\mathrm{Fr.}$ were calculated by (\ref{eq_RTMPC Y error}) and (\ref{eq_RTMPC Y error_Pro}), respectively:
\begin{subequations}
\begin{align}
e_\mathrm{Fr.}(k) &= g_\mathrm{Fr.}(x(k),u(k),\tilde{M}) \nonumber\\ 
&\quad\quad\quad\quad\quad - g_\mathrm{Fr.}(\bar{x}(k),u_\mathrm{MPC}(k),M),\\
e^*_\mathrm{Fr.}(k) &= g_\mathrm{Fr.}(x(k),u^*(k),\tilde{M}) \nonumber\\
&\quad\quad\quad\quad\quad- g_\mathrm{Fr.}(\bar{x}(k),u_\mathrm{MPC}(k),M),
\end{align}
\end{subequations}
and analyzed its upperbound that denotes the required tightening parameter $\gamma $, which is shown in Fig. \ref{fig_Set_govern}. Without the proposed governor, the tightening parameter $\gamma = 0.5$ was required, meaning that a very conservative MPC using only 50$\%$ of the friction limit should be imitated to satisfy the MPC's constraint for the changed model parameters in the target domain. On the other hand, with the proposed governor, $\gamma = 0.2$ was sufficient, allowing for a less conservative MPC to be imitated that utilizes 80$\%$ of the friction limit, significantly improving control performance.

Fig.~\ref{fig__govern2} presents the resulting control inputs (Fig.~\ref{fig_Input_govern}) and cart position (Fig.~\ref{fig_Posi_govern}) with and without the governor. As shown in Fig.~\ref{fig_Input_govern}, the governor increased the control input to compensate for the cart-pole's increased mass, demonstrating that the proposed framework can overcome the DNN's limitation of not adapting to model parameter changes. Therefore, with the governor, a less conservative MPC could be imitated, whereas without it, a highly conservative MPC was required, leading to an overly conservative DNN policy. Fig.~\ref{fig_Posi_govern} shows that the cart-pole reached the origin much faster when the governor was applied. The 2\% settling time was reduced by approximately 65\% (from 5.4s to 1.9s), which can be attributed to the robustness of the proposed framework under model parameter variations.

\subsection{Vehicle collision avoidance system}

\begin{figure}[t]
    \centering
    \includegraphics[width=1\columnwidth]{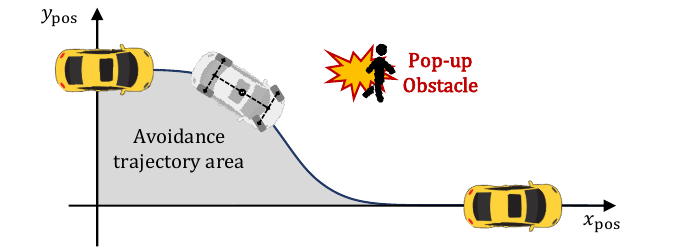}
    \caption{Vehicle collision avoidance system.}
    \label{fig_vehicle}
    \vspace{0mm}
\end{figure}

A practical application to a vehicle collision avoidance control case is also studied. The applied vehicle plant has 11 states and 5 inputs, including input lags in the lower-level controllers, making it a more complex system than the cart-pole control case. As shown in Fig. \ref{fig_vehicle}, the control objective is to regulate the lateral position of the vehicle to the safe lane ($y_\mathrm{pos}=0$) to avoid collision while preventing front tires' saturation to maintain vehicle maneuverability. To achieve this, an MPC was designed, followed by imitation learning, and applied to the proposed control framework. The compositions of the MPC, nominal prediction model, cost function, and constraints are briefly explained, as they are not the main focus of this paper. Details can be found in  \cite{kim2023imitation}.

The system dynamics for the vehicle can be given by:
\begin{subequations}\label{eq_Balance}
\begin{gather}
\dot{x}_{v} = f_{v} \left( x_{v},u_{v} \right),\\
x_{v}=[x_\mathrm{pos},y_{\mathrm{pos}},\psi,\dot{\psi},\beta,v_x ]^{T},\\ 
u_{v}= \left[ F_{x,fl}, F_{x,fr}, F_{x,rl}, F_{x,rr}, \delta \right]^{T},
\end{gather}
\end{subequations}
using a nonlinear full-car model with a brush tire model \cite{pacejka2005tire}. $x_{\mathrm{pos}}$ and $y_{\mathrm{pos}}$ are the global longitudinal and lateral positions of the vehicle; $\psi$ is the yaw angle; $\beta$ is the slip angle; $v_x$ is its longitudinal velocity; $\delta$ is the front steering angle. $F_{x}$ and $F_{y}$ are the longitudinal and lateral forces, respectively, acting on each tire. The subscripts $fl$, $fr$, $rl$, and $rr$ represent the front left, front right, rear left, and rear right tires, respectively. Also, input lags inherited in the lower controllers can be considered as a first-order linear system as follows:
\begin{subequations}\label{eq_Lowercontroller}
\begin{gather}
\tau_{F_{x}} \dot{F}_{x,j}  = - F_{x,j} + F_{x,j}^{in} , \text{ for } j=fl,fr,rl,rr,\\
\tau_{\delta} \dot{\delta}= - \delta + \delta^{in},
\end{gather}
\end{subequations}
\noindent where $\tau_{F_x}$ and $\tau_{\delta}$ are the time constants for the lower controllers, and $F_{x,j}^{in}$ and $\delta^{in}$ are the control variables.

The control objective is to regulate $y_{\mathrm{pos}}$ of the vehicle to the origin to avoid collision. To formulate the MPC's optimization problem, the prediction model was made, derived from (\ref{eq_Balance}), (\ref{eq_Lowercontroller}), and discretized using RK4 with sampling time 0.05s, as follows:
\begin{subequations}
\begin{align}
&x(k+1)= f(x(k),u(k),M),\\
&x = [x_v,u_v]^T,\\
&u = [F_{x,fl}^{in}, F_{x,fr}^{in}, F_{x,rl}^{in}, F_{x,rr}^{in}, \delta^{in}]^{T},
\end{align}
\end{subequations}
with vehicle parameters $M=[m,I_z,l_f,h,C_y]$ representing vehicle mass $m$, moment of inertia $I_z$, distance of the CG point from the front axle $l_f$, CG height $h$, and tire's normalized cornering stiffness $C_y$. A cost function for collision avoidance was structured, which maximizes the space between the vehicle and the obstacle by penalizing the avoidance trajectory area:
\begin{equation} \label{cost function}
\begin{aligned}
J(x(k),&U(k)) =\sum_{i=0}^{N-1}    \left(\right. q_{\mathrm{avoid}} \left(  x_{i|k} \right)  \\
& \quad\quad\quad\quad + u_{v,i|k}^T R_{1} u_{v,i|k} + \Delta u_{v,i|k}^T R_{2} \Delta u_{v,i|k} \left. \right),
\end{aligned}
\end{equation}
\noindent with the horizon of $N=70$, positive definite matrices $R_1$, $R_2$ for penalizing the control inputs, and $q_\mathrm{avoid}(\bullet)$ penalizing the avoidance trajectory area using mensuration by parts as follows:
\begin{equation}
q_{\mathrm{avoid}}\left(   x \right) = \left| y_{\mathrm{pos}} v_{x} \cos \left( \psi + \beta \right) T \right|.
\end{equation}
\noindent Finally, constraints were applied to ensure vehicle maneuverability by preventing the front tires' grip saturation:
\begin{equation} \label{ineq1}
\bar{\mathbb{X}} = \{\bar{x} \mid\frac{\sqrt{\left(F_{x,j}\right)^2+\left(F_{y,j}\right)^2}}{\mu F_{z,j}} -1 \leq-\gamma \},
\end{equation}
for $j=fl,fr$. Note that $0 \leq \gamma \leq 1$ is the tightening parameter. Additional constraints were added to keep braking forces negative, avoid exceeding tire grip, prevent overshoot, and enforce a terminal constraint to ensure $y_\mathrm{pos}$ converges to the origin (control stability). 

\begin{table}[t]
\begin{center}
\caption{Initial Conditions for Collecting $u_\mathrm{nom}$ Demonstrations.}
\label{tab_Initial}
\begin{tabular}{|c|c|c|c|}
\hline {\text { State }} & {\text { Range }} & {\text { State }} & {\text { Range}} \\
\hline  $x_{\mathrm{pos}} \, [\mathrm{m}]$ & 0 & $y_{\mathrm{pos}} \, [\mathrm{m}]$ & 2, 3, 4 \\
\hline $\psi \, [\mathrm{deg}]$ & -3, 0, 3 & $\dot{\psi} \, [\mathrm{deg/s}]$ & -2, 0, 25 \\
\hline $\beta \, [\mathrm{deg}]$ & -0.5, 0, 0.5 & $v_x \, [\mathrm{km/h}]$ & 70, 80, 90 \\
\hline $F_x \, [\mathrm{N}]$ & -1000, 0 & $\delta \, [\mathrm{deg}]$ & -1, 0, 1  \\
\hline
\end{tabular}
\end{center}
\end{table}

\subsubsection{Proposed control framework}

By imitation learning, the designed MPC was approximated to a fully connected DNN, with 5 hidden layers of 45 nodes each:
\begin{equation}
u_\mathrm{MPC}(\bullet) \,\rightarrow \, \pi_{\theta^*_{\mathcal{S_\mathrm{nom}}}}(\bullet).
\end{equation}
MPC demonstrations were collected from $\mathcal{S_\mathrm{nom}}$, starting from various initial states as Table \ref{tab_Initial}. $F_x$ and $\delta$ represent the initial driver inputs when the DNN intervenes to avoid collisions, in which the braking forces are distributed across the axles by a factor of $\eta$ via the proportional valve:
\begin{subequations}
\begin{gather}
F_{x,fl} = F_{x,fr} = F_x, \\
F_{x,rl} = F_{x,rr} = \eta F_x.
\end{gather}
\end{subequations}

Consequently, a total of 102,060 demonstrations were collected and used to train the DNN. The demonstrations were uniformly sampled within the vehicle's operating domain in the source domain. Data efficiency could be further improved by adopting active learning strategies, such as selectively collecting additional data near regions where constraint violations are likely to occur (e.g., when tire forces approach the friction limit), or in areas where the DNN exhibits higher prediction uncertainty \cite{judah2012active}. Nevertheless, since the primary focus of this study is on overcoming the sim-to-real gap, a simple grid-based sampling method was employed for the demonstration collection.

\begin{table}[t!]
\begin{center}
\caption{Nominal and Actual Vehicle Parameters.}
\label{tab_Parameter}
\renewcommand{\arraystretch}{1.25}
\begin{tabular}{|c|c|c|c|c|c|}
\hline & $m \, [\mathrm{kg}]$ & $I_z \, [\mathrm{kg}\cdot\mathrm{m}^2]$ & $l_f\, [\mathrm{m}]$ & $h\, [\mathrm{m}]$ & $C_y\, [\text{-}]$ \\
\hline  $M$ & 1830 & 3771 & 1.41 & 0.51 & 17.6 \\
\hline $\tilde{M}$  & $\pm$9.2$\%$ & $\pm$10.6$\%$ & $\pm$9.2$\%$ & $\pm$8.3$\%$ & $+$18$\%$ \\
\hline
\end{tabular}
\end{center}
\end{table}

The ancillary controller was designed as a full-state feedback controller, with an LQR gain $K$ obtained by solving DARE for the linear longitudinal and lateral vehicle model equations to stabilize the X and Y position errors. Under the proposed framework, the control input $u(k)$ was formulated as:
\begin{equation}
u(k) = \pi_{\theta^*_{\mathcal{S_\mathrm{nom}}}}(\bar{x}(k)) + K(x(k) - \bar{x}(k)).
\end{equation}

Finally, the input refinement governor (\ref{eq_Model-based filter}) leads to the following form regarding the front tires' grip constraints (\ref{ineq1}):
\begin{equation}\label{eq_filter_proposed}
g_{\mathrm{Grip},j}(x(k),\tilde{M})=g_{\mathrm{Grip},j}(x(k),M), \, \mathrm{for}\,\, j=fl,fr,
\end{equation}
\noindent with $g_{\mathrm{Grip},j}(\bullet)$ representing the front tires' grip defined as:
\begin{equation}
g_\mathrm{Grip,j} (x(k),M)= \left. \frac{\sqrt{(F_{x,j})^2 + (F_{y,j})^2}}{\mu F_{z,j}} \right|_{(k)}-1.
\end{equation}
\noindent 
The proposed governor can be interpreted as matching the front tires' grip of the plant vehicle to the values intended by the DNN, considering the changed vehicle parameters. The refined input $u^*$ satisfying (\ref{eq_filter_proposed}) can be obtained by matching the longitudinal and lateral normalized forces of each wheel:
\begin{equation}\label{eq_filter_concept}
h_\mathrm{Grip}(x(k),\tilde{M})=h_\mathrm{Grip}(\bar{x}(k),M),
\end{equation}
where $h_\mathrm{Grip}(\bullet)$ is defined as:
\begin{equation}
\begin{aligned}
h(&x(k),M) \\
& =\left. \left[ \frac{F_{x,fl}}{\mu F_{z,fl}}, \frac{F_{x,fr}}{\mu F_{z,fr}}, \frac{F_{x,rl}}{\mu F_{z,rl}}, \frac{F_{x,rr}}{\mu F_{z,rr}}, \frac{F_{y,f}}{\mu F_{z,f}} \right]^T \right|_{(k)},
\end{aligned}
\end{equation}
\noindent with $F_{y,f}/\mu F_{z,f}$ denoting the normalized lateral tire forces for the front axles, which share identical values for the left and right tires by the brush tire model. Similar to the cart-pole case study, the governor can be physically interpreted as refining the longitudinal and lateral tire forces proportionally to changes in the normal force; that is, if the vehicle mass increases, the braking forces of each wheel increase accordingly, and if the tire parameters changes (cornering stiffness), the steering angle input is refined matching the normalized lateral tire force.

However, due to the delay in the lower actuator (\ref{eq_Lowercontroller}), the refined input $u^*(k)$ at step $k$ is delayed and impacts $x(k+1)$ at step $k+1$, it is not possible to compute $u^*(k)$ at step $k$ to satisfy (\ref{eq_filter_concept}). Therefore, the governor is designed to compute $u^*(k)$ such that (\ref{eq_filter_concept}) is satisfied in the predicted states at step $k+1$:
\begin{equation}\label{eq_filter_real_pred_pure}
h_\mathrm{Grip}(x^*_{1|k},\tilde{M})=h_\mathrm{Grip}(\bar{x}_{1|k},M),
\end{equation}
\noindent where $x^*_{1|k}$ and $\bar{x}_{1|k}$ are the one-step predicted states, which can be obtained using the vehicle model as follows:
\begin{subequations}
\begin{gather}
x^*_{1|k} = f(x(k),u^*(k),\tilde{M}),  \\
\bar{x}_{1|k} = f(\bar{x}(k),u(k),M).
\end{gather}
\end{subequations}

\subsubsection{Results}

An E-class sedan was controlled in the CarSim target domain, from $y_\mathrm{pos} = 3.5$m to $y_\mathrm{pos}=0$m at 80km/h on a straight road with vehicle parameters $\tilde{M}$ set differently from the nominal parameters $M$ as Table \ref{tab_Parameter}. Note that the DNN was trained only on the nominal vehicle model with parameters $M$ in the source domain. On the other hand, CarSim, used as the target domain, is a high-fidelity vehicle dynamics simulator that includes tire rolling resistance, aerodynamic drag, suspension dynamics, and tire nonlinearities, resulting in significant sim-to-real gaps that are not accounted for in the source domain. It represents the real-world vehicle behaviors, providing a challenging validation setting for the proposed framework. 

\begin{figure}[t!]
    \centering
    \subfloat[Controlled vehicle position.]{%
        \includegraphics[width=1\columnwidth]{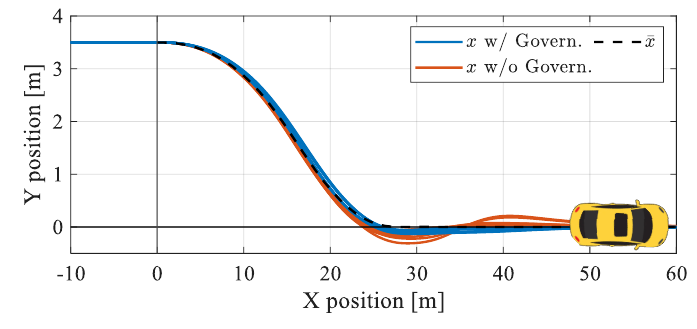} 
        \label{fig_Posi_govern_vehicle}
    }
    \vfill 
    \subfloat[Front tires' grip usage.]{%
        \includegraphics[width=1\columnwidth]{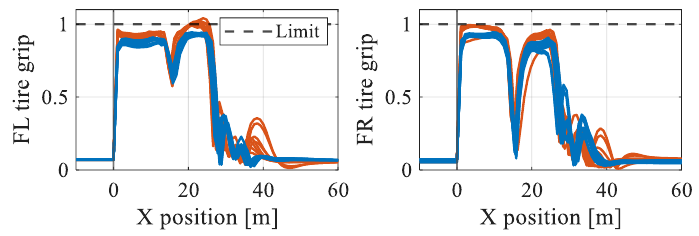} 
        \label{fig_grip_govern_vehicle}
    }
    \caption{Controlled vehicle position and front tires' grip usage with and without the governor.}
    \label{fig__govern2_vehicle}
\end{figure}

\begin{figure}[t!]
    \centering
    \includegraphics[width=1\columnwidth]{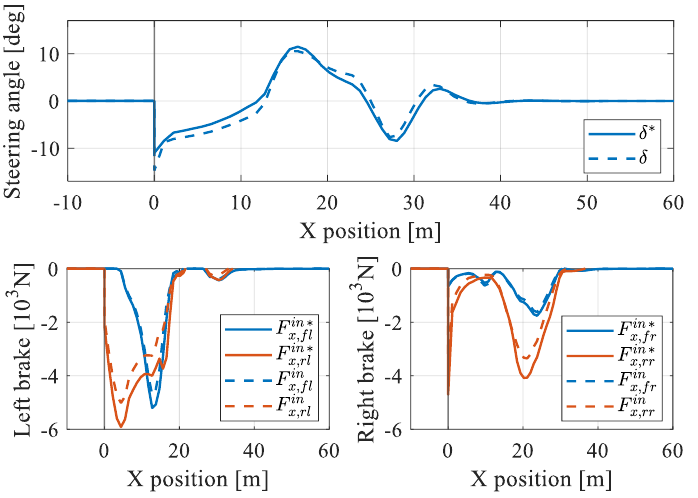} 
    \caption{Control inputs with (blue) and without (red) the governor.}
    \label{fig_Input_govern_vehicle}
    \vspace{0mm}
\end{figure}

Fig. \ref{fig__govern2_vehicle} shows the controlled vehicle position (Fig. \ref{fig_Posi_govern_vehicle}) and front tires' grip usage (Fig. \ref{fig_grip_govern_vehicle}) with and without the governor using the same DNN controller. The overlaid trajectories correspond to 32 combinations of model parameter variations $\tilde{M}$ within the range specified in Table~\ref{tab_Parameter}. Overall, the proposed controller effectively addressed the sim-to-real gap, guiding the vehicle to the safe lane as intended by the DNN, as shown in Fig. \ref{fig_Posi_govern_vehicle}. It is also noteworthy that, because the proposed framework enabled the DNN to remain within the source domain, simple behavior cloning was sufficient for implementation, without requiring more advanced imitation learning methods such as DAgger\cite{ross2011reduction}.

\begin{figure}[t!]
    \centering
    \includegraphics[width=1\columnwidth]{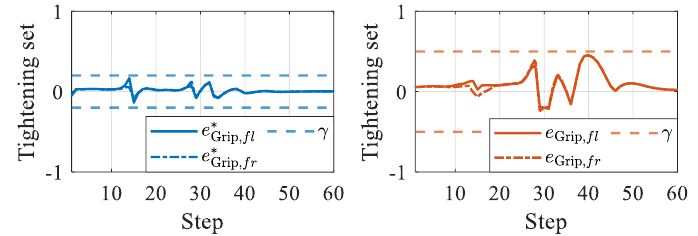} 
    \caption{Required tightening parameter $\gamma$ with (blue) and without (red) the governor.}
    \label{fig_Set_govern_vehicle}
    \vspace{0mm}
\end{figure}

The controller without the governor generated control inputs suitable only for the nominal model parameters and failed to adapt to the changed vehicle parameters. Consequently, the front tires' grip exceeded the limit, leading to tire saturation and unstable maneuvers during the evasion, as shown in Fig.~\ref{fig_grip_govern_vehicle}. These results confirm that, without the governor, more conservative constraint tightening is required to cope with parameter variations in the target domain. On the other hand, the controller with the governor refined the control inputs to adapt to the changed vehicle parameters. The maximum tire grip usage was consistently maintained around 90\%, and the evasive trajectory remained closer to the nominal state, regardless of the parameter variations. This indicates that the vehicle was more effectively controlled as intended by the DNN, even under different model parameters.  

Fig.~\ref{fig_Input_govern_vehicle} presents the refined control inputs generated by the governor under a specific parameter variation, where all parameters were increased within the range specified in Table~\ref{tab_Parameter}. The braking forces were increased to compensate for the higher mass and inertia (\(m, I_z\)) and for the changes in the center-of-gravity positions (\(l_f, h\)), while the steering angles were slightly reduced to account for the increased cornering stiffness (\(C_y\)). As a result, the grip constraints were satisfied, ensuring stable vehicle maneuverability during the evasion maneuver.

It was confirmed that the input refinement governor did indeed reduce the tightening parameter. Under the same parameter variation, Fig.~\ref{fig_Set_govern_vehicle} shows the sequences of \(e^*_{\mathrm{Grip},j}(k)\) and \(e_{\mathrm{Grip},j}(k)\) for \(j = fl, fr\), calculated using (\ref{eq_RTMPC Y error}) and (\ref{eq_RTMPC Y error_Pro}), respectively. As mentioned, the tightening parameter should be at least as big as each absolute value. With the governor, a tightening parameter of \(\gamma = 0.1\) was sufficient to satisfy the tire grip constraints, fully utilizing the vehicle's handling capability by using 90\% of the front tires' grip. In contrast, without the governor, a larger \(\gamma = 0.5\) was required, meaning that a very conservative MPC, using only 50\% of the available grip, should be imitated to satisfy the constraints under the changed vehicle parameters in the target domain, significantly degrading avoidance performance. Consequently, even with the same DNN controller, applying the governor reduced overshoot beyond the safe lane by 60\% (from 0.31m to 0.12m), resulting in a more robust vehicle avoidance maneuver under model parameter variations.

Overall, the framework achieved a computation time of 1.1ms on a PC equipped with an Intel Core i7-10900 CPU and 32GB of RAM, demonstrating a significant improvement in computational efficiency compared to the original MPC, which required approximately 600s per step.


\section{CONCLUSION AND FUTURE WORK}

This study proposed a novel control framework to address the sim-to-real gap when applying imitation learning to MPC. Inspired by RTMPC, the DNN is composed of a nominal controller operating in the same environment as the source domain. Compared to DR baselines, the proposed method achieved better data efficiency. Additionally, an input refinement governor was introduced to refine the control input based on changes in model parameters, allowing a less conservative MPC to be imitated. 

In this study, two case studies of applying the proposed framework were shown by a cart-pole and a vehicle collision avoidance system example, offering a pathway for generalizing the proposed framework to other control problems. The proposed governor is in a highly constrained form, making it still challenging to apply to general control cases. Therefore, for future work, we plan to relax the governor form and validate it in wider control applications for high-dimensional systems, such as multi-vehicle scenarios, along with real-world validations.

\section*{DECLARATIONS}

\subsection*{Conflict of Interest}
The authors declare that there are no competing financial interests or personal relationships that could have influenced the work reported in this paper.

\subsection*{Authors' Contributions}
Seungtaek Kim was responsible for conceptualization, validation, and writing the original draft. Jonghyup Lee was responsible for reviewing and editing the final manuscript. Kyoungseok Han was responsible for project administration. Seibum B. Choi was responsible for supervision.

\subsection*{Funding } 
This work was supported by the BK21 FOUR Program of the National Research Foundation of Korea (NRF) grant funded by the Ministry of Education (MOE), the Korea government (MSIT), the Ministry of Trade, Industry, and Energy (MOTIE, Korea), and the Korea Evaluation Institute of Industrial Technology (KEIT). (No. RS-2024-00346702, No. 20023815, and No. 20018181)


\biography{Figure/bio_KST}{Seungtaek Kim}{received the B.S. degree in Mechanical Engineering from Yonsei University, Seoul, South Korea, in 2019, and the M.S. and Ph.D. degrees in Mechanical Engineering from Korea Advanced Institute of Science and Technology (KAIST), Daejeon, Korea, in 2021 and 2025, respectively. Since 2025, he has been working as a Postdoctoral Research Fellow at KAIST. His research interests include vehicle dynamics and control, control theory, and imitation learning.}

\biography{Figure/bio_LJH}{Jonghyup Lee}{received the B.S., M.S., and Ph.D. degrees in Mechanical Engineering from Korea Advanced Institute of Science and Technology (KAIST), Daejeon, Korea, in 2015, 2017, and 2023, respectively. From 2023 to 2024, he was with Hyundai Motor Company as a Senior Research Engineer. Since 2024, he has been with the Department of Mechanical Systems Engineering at Sookmyung Women's University, Seoul, Korea, as an Assistant Professor. His research interests include autonomous driving decision and control, vehicle dynamics and control, and control theory.}

\biography{Figure/bio_HKS}{Kyoungseok Han}{received his B.S. degree in Civil Engineering with a minor in Mechanical Engineering from Hanyang University, Seoul, South Korea, in 2013, followed by M.S. and Ph.D. degrees in Mechanical Engineering from KAIST, Daejeon, South Korea, in 2015 and 2018, respectively. He is currently an Associate Professor in the Department of Automotive Engineering at Hanyang University, Seoul, South Korea. Prior to this, he worked as a Postdoctoral Research Fellow at the University of Michigan from 2018 to 2020 and as an Associate Professor at Kyungpook National University, Daegu, South Korea, from March 2020 to August 2024. His research interests include autonomous vehicle modeling and control, energy-efficient control of electric vehicles, reinforcement learning, and optimal control theory and its applications.}

\biography{Figure/bio_CSB}{Seibum B. Choi}{received a B.S. in Mechanical Engineering from Seoul National University, Seoul, Korea, an M.S. in Mechanical Engineering from KAIST, Daejeon, Korea, and a Ph.D. in control from the University of California, Berkeley, CA, USA, in 1993. From 1993 to 1997, he was involved in developing automated vehicle control systems at the Institute of Transportation Studies, University of California. From 2006 to 2009, he was with TRW in Livonia, MI, USA, where he developed advanced vehicle control systems. Since 2009, he has been a faculty member in the Mechanical Engineering Department of KAIST, Korea. His research interests include fuel-saving technology, vehicle dynamics and control, and active safety systems. Prof. Choi is a Member of the American Society of Mechanical Engineers, the Society of Automotive Engineers, and the Korean Society of Automotive Engineers.}
\clearafterbiography
\relax 

\end{document}